\documentclass[reprint, superscriptaddress, amsmath, amssymb, aps, prb]{revtex4-1}

\usepackage{graphicx}
\usepackage{dcolumn}
\usepackage{bm}
\usepackage{color}

\begin{document}

\preprint{APS/123-QED}

\title{Non-equilibrium steady state phases of the interacting Aubry-Andre-Harper model}

\author{Yongchan Yoo}
\affiliation{
Department of Physics, Condensed Matter Theory Center, and Joint Quantum Institute,
University of Maryland, College Park, Maryland 20742, USA
}
\author{Junhyun Lee}
\affiliation{
Department of Physics, Condensed Matter Theory Center, and Joint Quantum Institute,
University of Maryland, College Park, Maryland 20742, USA
}
\author{Brian Swingle}
\affiliation{
Department of Physics, Condensed Matter Theory Center, Maryland Center for Fundamental Physics, and Joint Center for Quantum Information and Computer Science, University of Maryland, College Park, Maryland 20742, USA
}
\date{\today}

\begin{abstract}
Here we study the phase diagram of the Aubry-Andre-Harper model in the presence of strong interactions as the strength of the quasiperiodic potential is varied. Previous work has established the existence of many-body localized phase at large potential strength; here, we find a rich phase diagram in the delocalized regime characterized by spin transport and unusual correlations. We calculate the non-equilibrium steady states of a boundary-driven strongly interacting Aubry-Andre-Harper model by employing the time-evolving block decimation algorithm on matrix product density operators. From these steady states, we extract spin transport as a function of system size and quasiperiodic potential strength. This data shows spin transport going from superdiffusive to subdiffusive well before the localization transition; comparing to previous results, we also find that the transport transition is distinct from a transition observed in the speed of operator growth in the model. We also investigate the correlation structure of the steady state and find an unusual oscillation pattern for intermediate values of the potential strength. The unusual spin transport and quantum correlation structure suggest multiple dynamical phases between the much-studied thermal and many-body-localized phases.
\end{abstract}

\maketitle

\section{Introduction}

Developments in highly engineered quantum many-body systems have given unprecedented experimental access to controlled quantum many-body models.
For a variety of model systems, it is now possible to observe quantum many-body dynamics over times long compared to the microscopic energy scales in the Hamiltonian.
This influx of new experimental data has reinvigorated numerous conceptual questions in the foundations of the subject, including the conditions under which isolated quantum systems come to some kind of effective equilibrium state.

In this context, many-body localization (MBL)~\cite{BAA06, ZPP08, PH10, NH15, SPA15, PVP15, Imb16, HGP+17, BLS+17, KSH17, PV18, AL18, NRA18, AAB+19}, corresponding to a failure to reach equilibrium, has received intense attention for the insights it brings to the foundations of quantum statistical physics.
Many experimental investigations of localization physics in such systems have been carried out~\cite{REF+08, LPP+09, SHB+15, BLS+17, LBH+17, LBS+17, LSK+18, KSL+19}, with the non-equilibrium relaxation dynamics of the system typically being probed by preparing a suitable initial state and evolving it in time.
For example, if the system is initialized into an imbalanced state with more atoms on even than odd sites, monitoring the time-dynamics of the imbalance can reveal the onset of localization and the failure to thermalize.

From the experimental point of view, it has proven convenient to study localization in the context of so-called quasiperiodic systems~\cite{IOR+13, LLS+17, LKK+17, ZL18, XLH+19, DM19}.
These systems are typically formulated as lattice models subject to a periodic potential whose wavelength is incommensurate with the underlying lattice. When the imposed potential is strong enough, the system is driven to a localized phase. 

One important observation in systems with a quasiperiodic potential is anomalous transport characterized by a power-law in time spreading of the conserved particle number.
This power law can be super-diffusive, often as a result of proximity to integrability, or sub-diffusive. While the emergence of sub-diffusive transport in random systems can be explained by Griffiths effects, the correlated nature of the on-site potentials in a quasiperiodic system precludes conventional Griffiths effects.
There are a few theories to interpret the difference of the anomalous transport and localization in the quasiperiodic potential compared to the random potential~\cite{KSH17}, however, the physical origin is still unclear.
Here our interest will be transport physics of one-dimensional spinless fermions evolving according to the interacting Aubry-Andre-Harper (AAH) model~\cite{Har55, AA80}.

Prior work with this model has suggested that as the strength of the quasiperiodic potential is increased, the system can enter an intermediate regime of slow dynamics before the fully localized state is reached~\cite{KSL+19, LD20}.
A few theories have been proposed to explain these anomalous dynamics, including local fluctuations of energy density~\cite{LLA+15} and atypical transition rates between the single-particle eigenstates~\cite{LKK+17}. 
However, the origin of the slow dynamics is still under debate. 
Moreover, the observed dynamical behavior raises the question of the possible existence of intermediate many-body phases. 
Recently, such a slow intermediate phase was claimed based on a numerical observation that the butterfly velocity, the speed at which Heisenberg operators spread in a chaotic system, vanished well before the full localization transition~\cite{XLH+19}.

\begin{figure}[h]
\includegraphics[width=\linewidth]{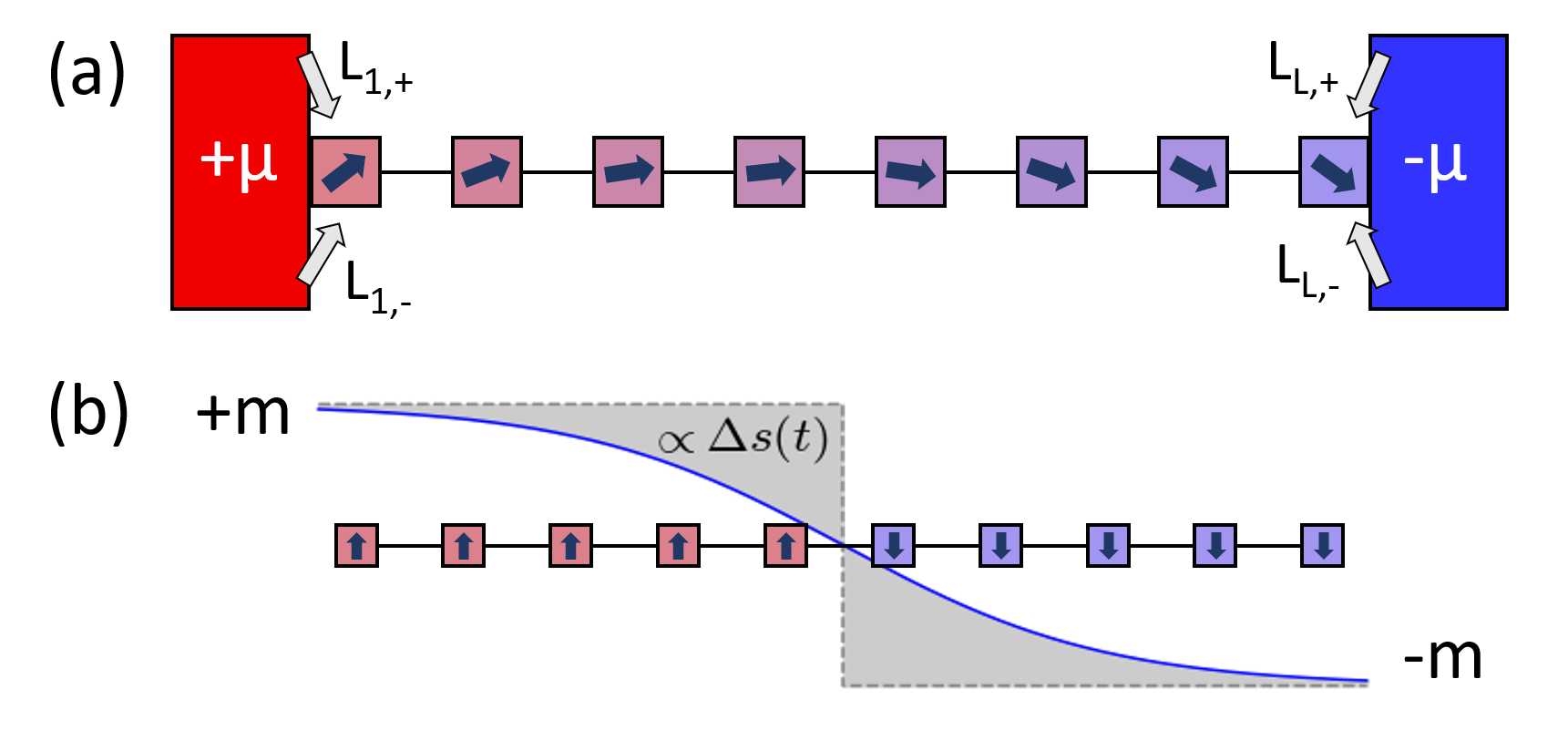}
\centering
\caption{(a) Lindbladian setting and (b) unitary setting for numerical simulations. The blocks with chemical potential $\pm \mu$ [(a)] are baths in contact with the boundaries of the spin chains via Lindblad operators, $L$. Blue solid line [(b)] represents the magnetization at time $t$ while the dotted lines are that on time $t = 0$. }
\label{fig:calcschem}
\end{figure}

In this work, we study transport and entanglement properties of non-equilibrium steady states (NESS) of the interacting AAH model in the strongly interacting regime under the influence of external baths coupled to the ends of the system [Fig.~\ref{fig:calcschem}(a)].
Using a standard fermion-to-spin mapping, the AAH model is converted into a corresponding spin model. In the spin language, the conserved quantity of interest is the spin/magnetization in the $z$-direction. 
Using a tensor network method built upon a matrix product operator (MPO) ansatz for the density matrix, the steady states of the model at large size can be obtained by evolving a Gorini-Kossakowski-Lindblad-Sudarshan (GKLS) master equation~\cite{Lin76, GK76} close to its dynamical fixed point.
By measuring the asymptotic scaling exponent characterizing magnetization transport, we show that there is a transition from superdiffusive transport to subdiffusive transport.

\begin{figure}[h]
\includegraphics[width=\linewidth]{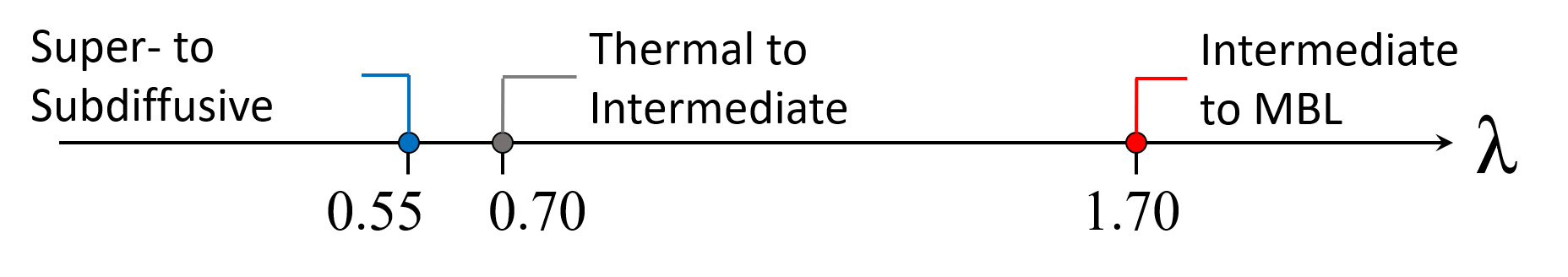}
\centering
\caption{Transport phase diagram of the interacting AAH model as a function of quasiperiodic potential strength $\lambda$.}
\label{fig:phasediag}
\end{figure}

The existence of this transition is further supported by a study of out-of-equilibrium magnetization domain wall dynamics using a different tensor network method based on unitarily evolving an initial state without any baths present [Fig.~\ref{fig:calcschem}(b)].
Interestingly, our data indicate that this transition before the point where the butterfly velocity vanishes.
Hence, we conclude that as the quasiperiodic potential strength is increased, the system first experiences subdiffusive magnetization transport then experiences subballistic operator growth, and then finally reaches a fully localized state.
These series of transition points are summarized in the phase diagram Fig.~\ref{fig:phasediag}.

The entanglement and correlation structure of the NESS is explored by studying the two-site quantum mutual information (QMI).
From the correlation pattern extracted from the QMI, we find that there are three distinctive regimes of potential strength, which are characterized by different decay trends in the QMI.
For small values of the potential, the QMI decays approximately monotonically.
As the strength increases, the system enters a regime showing strong modulation of the QMI with the same wavelength as the quasiperiodic potential.
Interestingly, this modulation suddenly disappears at even larger potential strength, providing another indication that the model supports very rich dynamics.

\section{Model and methods}

\subsection{Models}

We study the interacting AAH model with the following Hamiltonian~\cite{Har55, AA80}, 
\begin{equation}
H = J \sum_{i = 1}^{N - 1} \left( \sigma_{i}^{x} \sigma_{i + 1}^{x} + \sigma_{i}^{y} \sigma_{i + 1}^{y} + U \sigma_{i}^{z} \sigma_{i + 1}^{z} \right) + \sum_{i = 1}^{N} h_i \sigma_{i}^{z},
\end{equation}
where $\sigma_{i}$'s are the spin Pauli matrices at site $i$ and $U$ is the interaction strength. Units of time and energy are chosen such that $J = \hbar = 1$. The quasiperiodic potential is $h_i = 2 \lambda \cos(2 \pi \beta i + \phi)$, which is characterized by a strength $\lambda$ and an irrational wave-number $\beta$.

Without the interaction ($U = 0$), the AAH model experiences a transition from a ballistic state to a fully localized state at $\lambda = 1$~\cite{AA80}.
Thus, a distinguishing feature of the localization transition in the non-interacting case is that the system has a sharp transition at a single quasiperiodic potential strength. Previous studies~\cite{SDP17, ZL18, XLH+19, LD20} suggested that adding the interaction leads to various intermediate phases with slow dynamics.

It is notable that since Griffiths-type regions do not appear with the quasiperiodic potential~\cite{AAD+17, NRA18}, the physical origin of slow intermediate dynamics in the interacting AAH model should be distinct from models with random disorders. Here we study the physics of spin transport and correlation as a function of varying potential strength in the strong interaction regime, $U = 1.0$, with fixed irrational wave number $\beta = (\sqrt{5} - 1) / 2$ and global phase $\phi = 0$.

\subsection{Master equation and NESS}

Our primary tool for investigating the spin dynamics is to study non-equilibrium steady states of long chains driven at the boundaries by Markovian baths. The Markovian time evolution of the open system is described by the GKLS master equation~\cite{Lin76, GK76}:
\begin{equation}
\frac{d \rho}{dt} = \mathcal{L} (\rho) \equiv i [\rho, H] + \sum_{\nu} \left ( L_{\nu} \rho L_{\nu}^{\dagger} - \frac{1}{2} \left \{ L_{\nu}^{\dagger} L_{\nu}, \rho \right \} \right).
\end{equation}
Here, $\rho$ is the density matrix of the system and $L_{\nu}$'s are the Lindblad operators representing the baths with index $\nu$. 
We consider single-spin baths at the two boundaries of the system ($i = 1,\, N$) with a chemical potential bias $\mu$. 
The Lindblad operators then become $L_{1,\pm} = \sqrt{1 \pm \mu} \sigma_{1}^{\pm}$ and $L_{N, \pm} = \sqrt{1 \mp \mu} \sigma_{N}^{\pm}$.
Note that $\sigma^{\pm} = \frac{1}{2}(\sigma^{x} \pm i \sigma^{y})$. 
At infinite time, the solution of the GKLS equation can approach a steady state, and we attempt to access that steady state by evolving for a large but finite time. This distinction is important because, although we always evolve long enough for the state to be very slowly changing in time, the true infinite time steady state may nevertheless not be approximately reached, especially at larger values of the potential.

Technically, if the density matrix is mapped to a superket state $\lvert \rho \rangle \rangle$~\cite{ZV04}, the GKLS master equation takes a numerically practical form $\frac{d}{dt} \lvert \rho \rangle \rangle = \mathcal{L} \lvert \rho \rangle \rangle$ with the Liouvillian superoperator $\mathcal{L}$ as follows~\cite{MFS15}:
\begin{align}
\mathcal{L} = &-i H \otimes \mathbb{I} + i \mathbb{I} \otimes H^T \nonumber\\
			&+ \sum_{\nu} \left ( L_{\nu} \otimes (L_{\nu}^{\dagger})^T - \frac{1}{2} \left( L_{\nu}^{\dagger} L_{\nu} \otimes \mathbb{I} + \mathbb{I} \otimes (L_{\nu}^{\dagger} L_{\nu})^T \right) \right ).
\label{eq:L}
\end{align}
The NESS ($\rho_{\infty}$) is approximated by $\lvert \rho(t) \rangle \rangle = e^{\mathcal{L}t} \lvert \rho(0) \rangle \rangle$ for sufficiently large time $t$. It is known~\cite{PZ09} that the NESS calculated from Eq.~\eqref{eq:L} is independent of the choice of initial state unless $\rho_{\infty}$ and $\rho(0)$ have zero overlaps. 
We choose a product state $\lvert \rho(0) \rangle \rangle = \prod_i e^{-\mu_i \sigma_i^z}$ as the initial state with $\mu_i$s linearly interpolating between the chemical potential bias ($\pm \mu$) at the two ends of the chain. Small bias is enough to measure the spin transport in this model while not perturbing the system too much; the calculations below take $\mu = 0.01$. This choice of initial state also appears to aid in a quick convergence to a quasi-steady state.

\subsection{Numerical method: Tensor networks}

In the superket-superoperator formalism, the density matrix and Liouvillian naturally map to a matrix product state (MPS) and matrix product operator (MPO), respectively. 
We choose the spin Pauli matrices $\sigma^{\alpha}$ $(\alpha = 0,\,x,\,y,\,z)$ as a basis for the MPS and MPO ($\sigma^0$ is the identity matrix). For instance, the dissipative part of the Liouvillian superoperator has a simple $4 \times 4$ matrix representation in this basis.
We then apply the time-evolving block decimation (TEBD) method~\cite{Vid03, Vid04} to the Liouvillian superoperator equation ($\lvert \rho(t) \rangle \rangle = e^{\mathcal{L}t} \lvert \rho(0) \rangle \rangle$  and Eq.~\eqref{eq:L}).
First, we decompose the propagator $e^{\mathcal{L}t}$ into small time-steps, $e^{\mathcal{L} \Delta t}$, and also write the Liouvillian superoperator as $\mathcal{L} = \mathcal{L}_1 + \mathcal{L}_2$ using the Suzuki-Trotter decomposition~\cite{Suz93}, where $\mathcal{L}_1$ and $\mathcal{L}_2$ are sums of mutually commuting terms. 
The use of the Suzuki-Trotter decomposition is justified as the Liouvillian (both the Hamiltonian and the dissipative term) have at most nearest-neighbor couplings. We use the second-order Suzuki-Trotter decomposition and time steps as small as $\Delta t = 0.05$ in our numerical simulations.

For systems with large quasiperiodic potential (typically $\lambda \sim 1.5$), the relaxation time is very long due to the slow dynamics.
To converge efficiently to the NESS in these cases, we choose different simulation parameters for the early and later stages of the time evolution, with the expectation that the intermediate time dynamics becomes irrelevant as the steady state is approached. At the early stage, a relatively small maximum bond dimension ($\chi =  32$) and a large time step ($\Delta t > 0.1$) are used to quickly drive the system into the rough neighborhood of the NESS. Then, during the remainder of the simulation, we carefully approach the NESS with a larger maximum bond dimension ($\chi = 128$) and a smaller time step ($\Delta t = 0.05$).

Expectation values of local observables ($\mathcal O$) are calculated as usual, $\langle \mathcal O \rangle = \text{tr}(\mathcal O \rho_\infty) / \text{tr}(\rho_\infty)$.
However, fluctuations are inevitably present in expectation values calculated from our approximate NESS, depending on the convergence. 
Therefore, throughout our numerical studies, we average over the expectation values of many successive Suzuki-Trotter time steps to effectively reduce the effect of fluctuations. For each observation, we monitor the convergence of the time-step averaged value and use an appropriate number of time-steps so that the result is converged. The number of sampling steps typically varies from 50 to 1000, depending on the simulation parameters of the model.

\section{Spin transport}

\begin{figure}[t]
    \includegraphics[width=\linewidth]{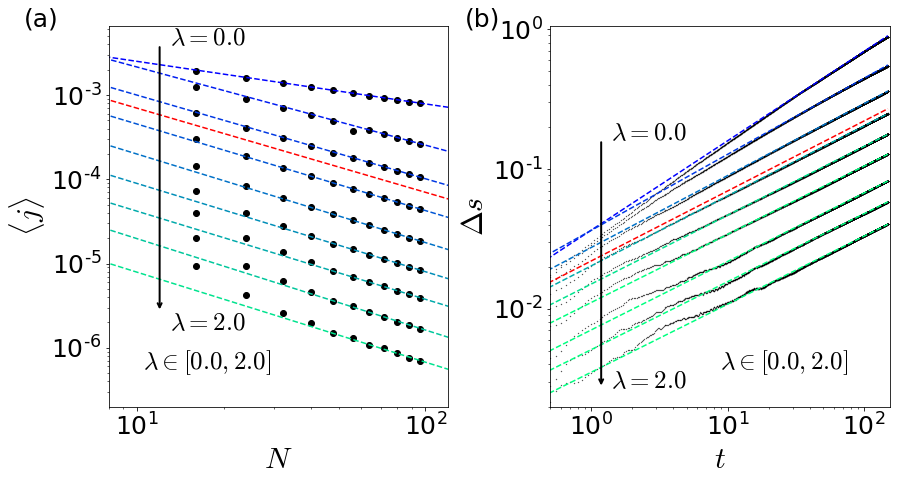}
    \centering
    \caption{The spin transport of the interacting AAH model. (a) The average spin current $\langle j \rangle$ of the NESS for different quasiperiodic potential strengths as a function of the system size $N$. The dashed lines are the best asymptotic inverse power-law fittings. (b) The spread portion of the total magnetization $\Delta s$ for different quasiperiodic potential strengths as a function of the time $t$. The gray dashed lines are the best asymptotic power-law fittings. The red dashed lines describe the diffusive transport scalings.}
    \label{fig:jands}
\end{figure}

To observe different dynamical phases of the AAH model, we concentrate on the average spin current of the system in NESS as a function of the quasiperiodic potential strength $\lambda$.
The spin current operator for the $i$-th bond is $j_i = 2 (\sigma_{i}^{x}\sigma_{i + 1}^{y} - \sigma_{i}^{y}\sigma_{i + 1}^{x})$; its expectation reaches an $i$-independent value (per the continuity equation) as the system approaches the NESS.

Diffusive transport is characterized by Fick's law, $\langle j \rangle= - D \, \partial_i \langle \sigma_i^z \rangle$, where $D$ is the diffusion constant. The derivative can be approximated as $\frac{\langle \sigma_N^z \rangle - \langle \sigma_1^z \rangle}{N}$ in the NESS. This length dependence of $\langle j \rangle$ can be generalized to non-diffusive situations using a scaling exponent $\gamma$:
\begin{align}
	\langle j \rangle = - D \frac{\langle \sigma_N^z \rangle - \langle \sigma_1^z \rangle}{N^\gamma}.
\end{align}
$\gamma = 1$ corresponds to Fick's law of diffusive transport while $\gamma = 0$ indicates ballistic transport.
Superdiffusive and subdiffusive transport correspond to $\gamma < 1$ and $\gamma > 1$, respectively. 

In the boundary-driven spin chain we consider [Fig.~\ref{fig:calcschem} (a)], the boundary magnetization is constrained via the chemical potential imbalance $\langle \sigma_N^z \rangle - \langle \sigma_1^z \rangle \approx - 2 \mu$.
Thus $\langle j \rangle \sim 1/N^\gamma$, and we can determine the exponent $\gamma$ by directly observing how the spin current scales with the system size $N$. This analysis is shown in Fig.~\ref{fig:jands}(a) which plots $\langle j \rangle$ as a function of $N$ in logarithmic scale. A linear fit, where the magnitude of the slope gives $\gamma$, is obtained from a series of large $N$ values ($N \geq 72$) to reduce finite size effects and to include the effect of self-averaging of $\phi$. The increasing trend of $\gamma$ with $\lambda$ shows that the system experiences a transition from superdiffusive to subdiffusive transport as the quasiperiodic potential strength increases. 

To further validate this result, we use a second method discussed in Ref.~\cite{ZL18} to investigate the spin transport. In this approach, we eliminate the external baths at the boundary and observe the unitary evolution of a sharp domain wall. The initial state is $\lvert \rho(0) \rangle \rangle \propto \prod_{i = 1}^{N/2} e^{m \sigma_i^z} \otimes \prod_{i =N/2+1}^{N} e^{-m \sigma_i^z}$ (with appropriate normalization), where the spins are weakly polarized ($m = \frac{\pi}{1800}$) with a domain wall at the center of the system [Fig.~\ref{fig:calcschem} (b)]. Since the spreading of the domain wall is monitored throughout the whole time evolution (instead of only at very large times), we can reach a bigger system size within the unitary evolution set up ($N = 128$ and $\chi = 128$). 

The spreading of the domain wall can be quantified by the difference of magnetization from the initial state, 
\begin{equation}
\Delta s (t) \equiv 1 - \frac{1}{mN}\left ( \sum_{i = 1}^{N/2} \langle \sigma_i^z (t)\rangle - \sum_{i = N/2 + 1}^{N} \langle \sigma_i^z (t)\rangle \right),
\end{equation}
where $\Delta s (0) = 0$ and $\Delta s (t) \rightarrow 1$ as $t \rightarrow \infty$, for typical diffusive systems. We define the scaling exponent $\alpha$ via $\Delta s(t) \sim t^\alpha$.

The two scaling exponents $\gamma$ and $\alpha$ are related to each other by a relation $\gamma = \frac{1}{\alpha} - 1$~\cite{LZP17}, which is obtained by dimensional analysis of the spin current. 
Therefore the values of $\alpha$ corresponding to diffusive and ballistic transport are $\alpha = 1/2$ and $\alpha = 1$, respectively. 
The system is subdiffusive for $0 < \alpha < 1/2$ and superdiffusive when $1/2 < \alpha < 1$. Fig.~\ref{fig:jands}(b) shows the time-evolution of $\Delta s(t)$ for different quasiperiodic potential strengths ($\lambda$) with the scaling exponents ($\alpha$) estimated from a least-squares best fit.

\begin{figure}[t]
\includegraphics[width=0.9\linewidth]{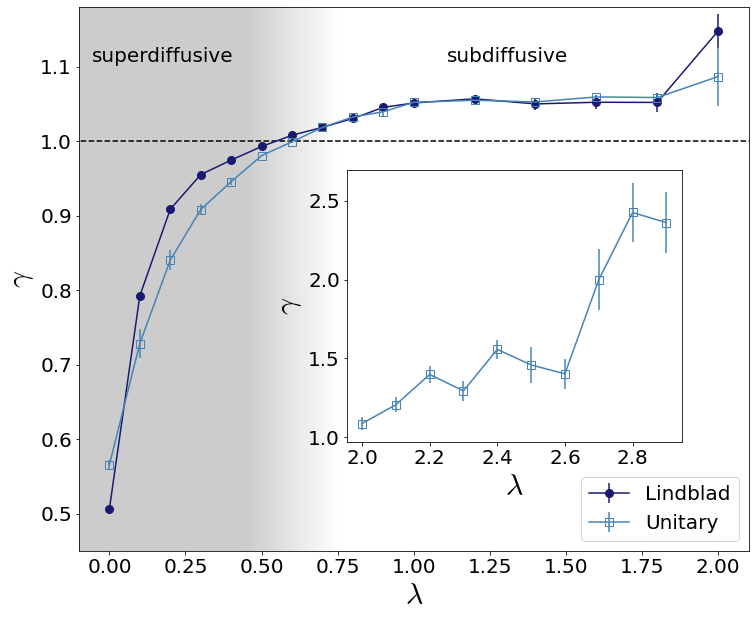}
\centering
\caption{The scaling exponent $\gamma$ from the Lindbladian and the unitary evolution as a function of the quasiperiodic potential strength. The (non-)shaded area represents the (subdiffusive) superdiffusive regime. The inset figure shows the $\gamma$ for $\lambda \geq 2.0$ from the unitary evolution. The error bar indicates the standard deviation from the fitting samples.}
\label{fig:gamma}
\end{figure}

Fig.~\ref{fig:gamma} compares the two methods of extracting $\gamma$ and shows that the results are largely consistent, especially in the subdiffusive regime. Without the quasiperiodic potential ($\lambda = 0$), the interacting AAH model reduces to the isotropic Heisenberg XXZ model and approximately gives the expected result of $\gamma = 0.5$~\cite{Zni11}. Starting from this superdiffusive regime, $\gamma$ increases with $\lambda$ above the critical value for exact diffusion ($\gamma = 1$), demonstrating a dynamical phase transition from superdiffusive to subdiffusive transport. The critical point is around $\lambda_c \approx 0.55 \,(0.60)$ according to the Lindbladian (unitary) dynamics. Note the difference from the XXZ model with the uniform random disorder, where the change in $\gamma$ is discontinuous and has a diffusive \emph{phase}~\cite{ZSV16}. By contrast, at least for the sizes and times probed in this interacting AAH model, $\gamma$ changes continuously in the superdiffusive regime and the exact diffusion only occurs at a \emph{point}. We originally expected a diffusive regime for weak potential strength but did not observe it in the system sizes studied.

A previous study on the interacting Aubry-Andre model with fermions revealed a critical point $\lambda_B$ between the thermal phase and a new intermediate phase~\cite{XLH+19}. 
This intermediate ``S phase'' is characterized by vanishing butterfly velocity and a power-law effective lightcone.
Interestingly, our best estimate is that the transport transition ($\lambda_c \approx 0.55$) occurs before the onset of the S phase ($\lambda_B \approx 0.7$) for identical parameters. This suggests that there are two subphases characterized by distinct spin transport physics within the thermal phase: (i) thermal and superdiffusive phase (ii) thermal and subdiffusive phase. 

Ref.~\cite{XLH+19} also showed the transition from S phase to an MBL phase occurs at $1.7 < \lambda_{\text{MBL}} < 1.9$.
In our boundary driven system, we calculate up to $\lambda = 2.0$ and observe the dynamics become extremely slow from around $\lambda \approx 1.7$.
Although this significant increase in relaxation time is a sign of the possible MBL transition, it also makes it nearly impossible to reach the NESS in this regime.
One notable point is that $\gamma$ exhibits a plateau-like behavior staying near the same value for $1.0 < \lambda < 2.0$.

For $\lambda > 2.0$, $\gamma$ dramatically increases with increasing $\lambda$ as shown in the inset of Fig.~\ref{fig:gamma}. The large error bars at those $\lambda$ values are due to the very slow dynamics which results in almost no changes of the value of $\Delta s(t)$. Accurately locating the localization transition is difficult with such observables, however, similar trends observed~\cite{IRR+16, LLS+17, DM19} in other `imbalance' parameters are indicative of a transition.

Incorporating our results with the previous work~\cite{XLH+19}, an updated phase diagram for the strongly interacting AAH model is shown in Fig.~\ref{fig:phasediag}. The phase diagram consists of four dynamical phases: (i) thermal and superdiffusive phase ($\lambda < \lambda_c$); (ii) thermal and subdiffusive phase ($\lambda_c < \lambda < \lambda_B$); (iii) subdiffusive S phase ($\lambda_B < \lambda < \lambda_{\text{MBL}}$); and (iv) MBL phase ($\lambda_{\text{MBL}}<\lambda )$.

\section{Correlation structure}

\begin{figure}[t]
\includegraphics[width=\linewidth]{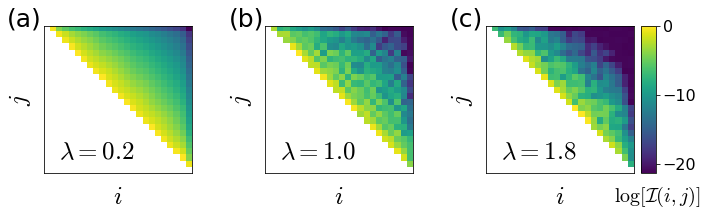}
\centering
\caption{The colomap matrix representation of logarithm of the renormalized two-site quantum mutual information of the NESS of the interacting AAH model, $\log \mathcal{I}(i, j)$, for system size $N = 24$ and different quasiperiodic potential strength $\lambda$; (a) $\lambda = 0.2$; (b) $\lambda = 1.0$; (c) $\lambda = 1.8$.}
\label{fig:colormap}
\end{figure}

We further examine the correlation and entanglement structure of the NESS of the interacting AAH model. The initial motivation for this calculation was to check that the low entanglement assumption of TEBD is valid for the boundary driven NESS; we subsequently uncovered an interesting pattern in the quantum correlations. 

These correlations, which include both classical effects and entanglement, can be quantified using the quantum mutual information (QMI), which has also been studied in the context of metal-insulator transition in the noninteracting AAH model and the interacting disordered Hubbard chain~\cite{DTBB+17}. The QMI $\mathcal{I}(A, B)$ for two subsystems $A$ and $B$ is given by the following formula~\cite{GPW05}:
\begin{align}
\mathcal{I}(A, B) = S(A) + S(B) - S(A \cup B),
\end{align}
where $S(A) = - \text{tr}(\rho_A \log \rho_A)$ is the von Neumann entropy of $A$ with reduced density matrix $\rho_A$. In particular, we consider the two-site QMI $\mathcal{I}(i, j)$, where the two subsystems are the $i$-th and $j$-th ($i \neq j$) sites of the system. 
We calculate QMI from the partial trace of the NESS obtained with the MPO Lindbladian evolution for each parameter.

Fig.~\ref{fig:colormap} shows $\mathcal{I}(i, j)$ for the interacting AAH model for several values of $\lambda$ in a logarithmic colormap scale.   
For small quasiperiodic potential strengths [Fig.~\ref{fig:colormap}(a)] the two-site QMI decreases monotonically and smoothly as $x = |i-j|$ increases. 
On the other hand, although the overall decay trend persists, non-monotonicity appears in $\mathcal{I}(i, j)$ for larger values of $\lambda$, noticeable by the checkerboard pattern [Fig.~\ref{fig:colormap}(b)]. 
As we further increase $\lambda$ [Fig.~\ref{fig:colormap}(c)], one observes the non-monotonicity effect decrease again.

\begin{figure}[t]
\includegraphics[width=0.9\linewidth]{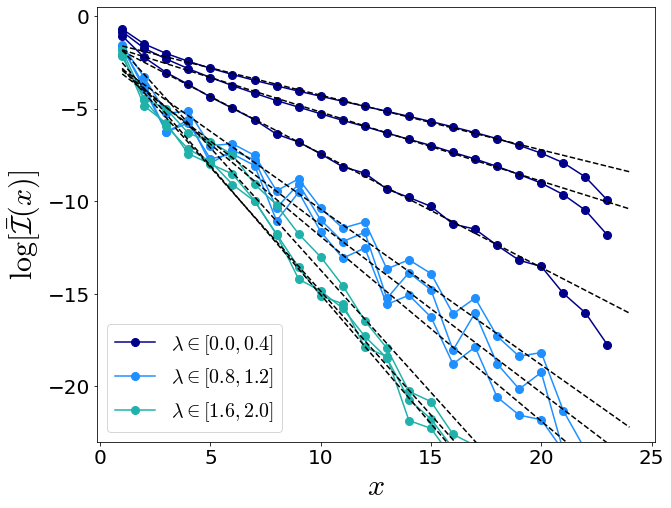}
\centering
\caption{The plot of the logarithm of the spatial and Trotter-time averaged two-site quantum mutual information of the NESS of the interacting AAH model, $\log \bar{\mathcal{I}} (x)$, as a function of the distance between two sampling sites $x$. The three distinct groups are highlighted separately. The dashed gray lines are the best exponential fittings for each $\lambda$.}
\label{fig:logI}
\end{figure}

We quantify this ``non-monotonicity'' pattern by introducing the averaged QMI $\bar{\mathcal{I}}(x) \equiv (1/N_{x})\sum_{|i-j| = x} \mathcal{I}(i,j)$, where $N_x$ is the number of ${i,j}$ combinations satisfying $|i-j| = x$.
Fig.~\ref{fig:logI} plots $\bar{\mathcal{I}}(x)$ in log-scale and the two-site correlation length can be read from the inverse slope. 
We clearly see the three distinct regimes of $\lambda$ as in Fig.~\ref{fig:colormap}: (i) smooth and monotonic decrease for $\lambda < 0.5$, (ii) large oscillations for $0.5 < \lambda < 1.4$, and (iii) suppressed oscillations for $1.4 < \lambda $. (Note that the values of $\lambda$ here are approximate and do not represent sharp critical values.)
The first transition from (i) to (ii) seems to be a crossover as the oscillation builds up continuously, while the vanishing of the oscillation in (ii) to (iii) is more abrupt pointing towards a phase transition. 
The correlation length decreases rapidly in regime (i) while the decrease slows down in (ii) and almost saturates in (iii). 

\begin{figure}[t]
\includegraphics[width=0.9\linewidth]{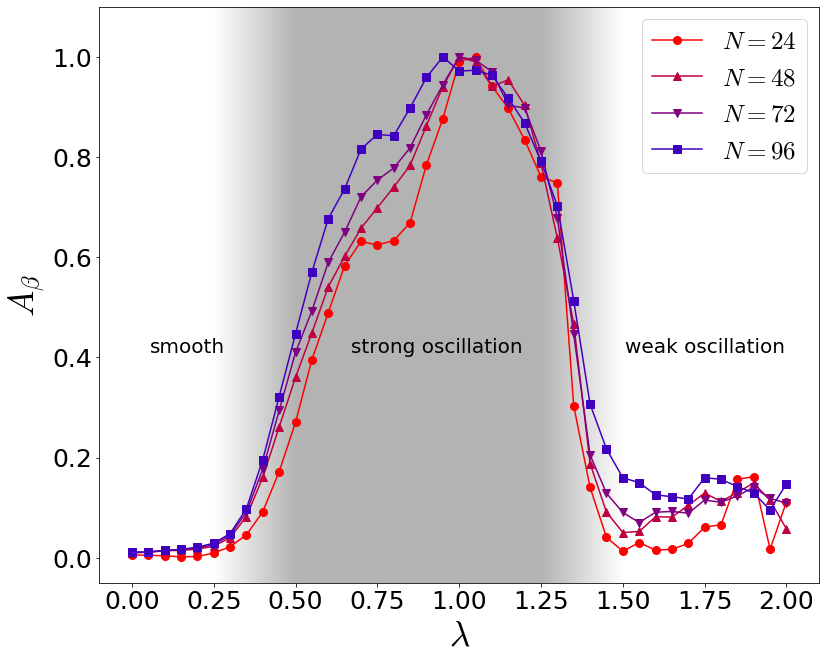}
\centering
\caption{The normalized amplitude of the modulation $A_{\beta} (\lambda, N)$ at $\omega = \beta$ as a function of the quasiperiodic potential strength for several system sizes.}
\label{fig:amps}
\end{figure}

The modulation in regime (ii) has the same wavenumber from the quasiperiodic potential ($\beta = (\sqrt{5} - 1)/2$). 
We extract the Fourier amplitude of this wavenumber ($A_\beta  = \int \bar{\mathcal{I}}(x) \cos (2\pi \beta x) dx$) and plot it as a function of $\lambda$ in Fig.~\ref{fig:amps}.
We confirm that $A_\beta$ smoothly increases near the first crossover point and steeply falls at the second crossover point. 
The data is consistent with the three distinct regimes of $\lambda$ based on the qualitative behavior of two-point QMI. The crossover points do not appear to coincide with the transport critical value $\lambda_c$, the critical point in Ref.~\cite{XLH+19} $\lambda_B$, or $\lambda_{\text{MBL}}$. 

This may be a convergence issue -- small $N$ suffers from finite size effects and large $N$ is less converged to the NESS. 
As in all tensor network calculations, there is also the possibility that the entanglement may be more severely affected by truncation compared to local observables. Note that the magnetization profile of the spin chain also has a modulation pattern with wavenumber $\beta$, however, it does not show any significant transition as we tune $\lambda$.

\begin{figure}[h]
\includegraphics[width=0.9\linewidth]{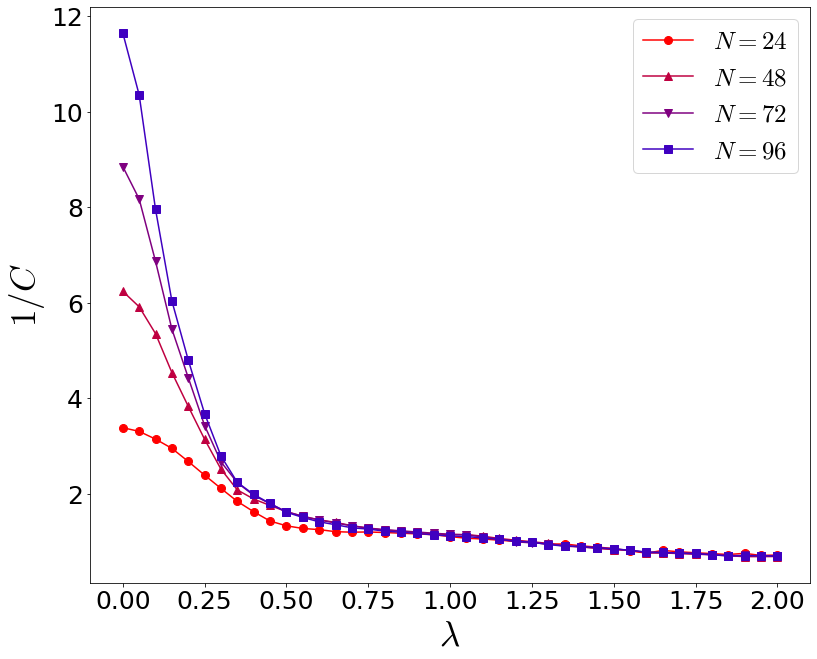}
\centering
\caption{The inverse of the slope $C$ of the linear fitting of $\log\left[\bar{\mathcal{I}}(x)\right]$ as a function of the quasiperiodic potential strength for several system sizes.}
\label{fig:corrlen}
\end{figure}

In addition, our expectation is that correlations in the NESS as measured by the QMI will exhibit an overall exponential decay with distance. However, the precise functional form of the QMI is non-trivial, as shown in Fig.~\ref{fig:logI}. A linear fit to $\log{ \bar{\mathcal{I}}}$ gives one measure of the correlation length; the results of this fit are shown in Fig.~\ref{fig:corrlen}. One sees a clear dependence on system size up until approximately $\lambda=0.75$ and weak dependence only on $\lambda$ thereafter. However, we know from Fig.~\ref{fig:amps} that the oscillations in the QMI continue to evolve with $\lambda$ up until approximately $\lambda=1.4$. Thus, it is only after the oscillations cease that the QMI profile becomes approximately independent of both system size and $\lambda$. It is interesting to note that $\lambda=0.75$ is close to the transition into the slow phase identified in~\cite{XLH+19}.

\section{Discussion}

In this work, using a combination of unitary and open system tensor network methods, we found a transition in spin transport from superdiffusive to subdiffusive as a function of increasing quasiperiodic potential strength. Our results, when combined with those of Ref.~\cite{XLH+19} which argued for a transition in the speed of operator growth, lead to a rich proposed phase diagram for the strongly interacting AAH model. In addition, we showed that the decay of the QMI also exhibits several distinct regimes as a function of the quasiperiodic potential strength, although these regimes may be separated by crossovers rather than genuine phase transitions.

This rich set of phenomenology deserves further study, especially since a simple physical picture is currently lacking. There are various proposals to explain the slow dynamics at intermediate potential strength, but it is not clear if any of these are sufficient. One idea is that, at a mean-field or Hartree-Fock-like level, the strong interaction significantly renormalizes the single-particle potential leading to a regime in which the single-particle spectrum contains a mixture of localized and delocalized states~\cite{XLH+19}. However, certainly scattering and other interaction effects need to be included.

To better understand the physics in the intermediate regime, it would be interesting to study energy transport in addition to spin transport. One idea is to make more precise the mean-field picture described above, perhaps connecting to prior discussions of mobility edges~\cite{MPH+15, BLS15, LPD+16, NG17, LD20}. A solvable model, perhaps using large $N$ technology~\cite{AGM+00}, might also be helpful in developing an analytic understanding of the physics. Our result that the system goes subdiffusive before the speed of operator growth vanishes seems consistent with recently proposed bounds on diffusivity in terms of the operator growth speed, but this connection is not sharp until we understand how to estimate the other timescales involved in the bound~\cite{HHM17, Luc17}. Finally, one could explore the physics discussed here in new regimes, for example, as a function of temperature.

\section*{Acknowledgments}
We thank C. White and S. Xu for helpful discussions.
B.S. is supported by DOE, ASCR, and QOALAS team.
Y.Y. and J.L. are supported by NSF-PFC at the JQI.
Y.Y. is also supported by Kwanjeong Educational Foundation.

\bibliographystyle{apsrev4-2}
\bibliography{NESSbib}


\appendix

\section{Entanglement growth}

The two time evolution methods we used have a small discrepancy in their asymptotic scaling exponent $\gamma$, in particular, in the regime of weak quasiperiodic potential strength [Fig.~\ref{fig:gamma}]. We investigate this difference mainly using the bipartite MPO entanglement entropy $\mathcal{S}$,~\cite{Zan01} which is defined as
\begin{equation}
    \mathcal{S} = - \sum_{i = 1}^{\chi} \lambda_{i}^2 \log{(\lambda_{i}^2)}.
\end{equation}
Here $\lambda_i$ is the normalized Schmidt values ($\sum_i \lambda_{i}^2 = 1$) of the matrix product density operator (MPDO) in the canonical form at the center of the system and $\chi$ is the dimension of the Schmidt values at the center.
Basically, the dynamics of the MPO entanglement entropy describes the growth of the entanglement in the MPO basis and also estimates the required maximum bond dimension for efficient MPDO simulation.

\begin{figure}[h]
\includegraphics[width=0.9\linewidth]{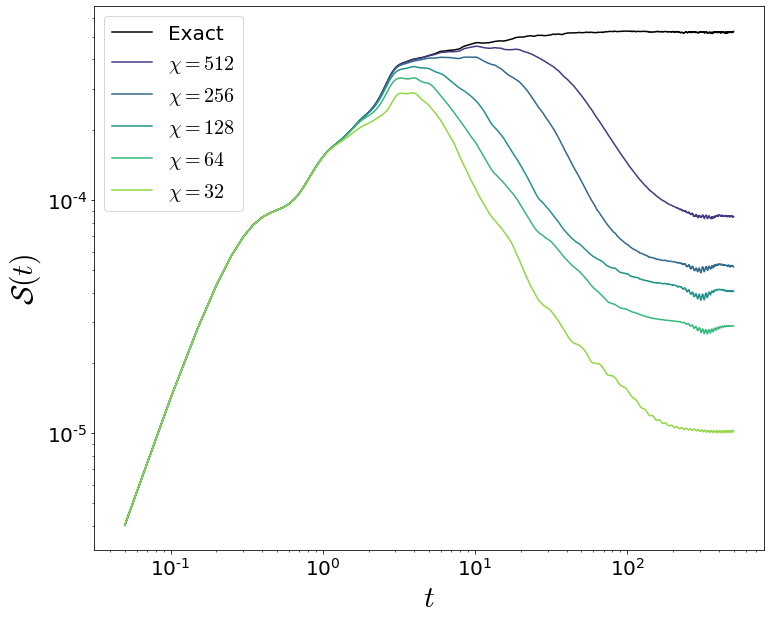}
\includegraphics[width=0.9\linewidth]{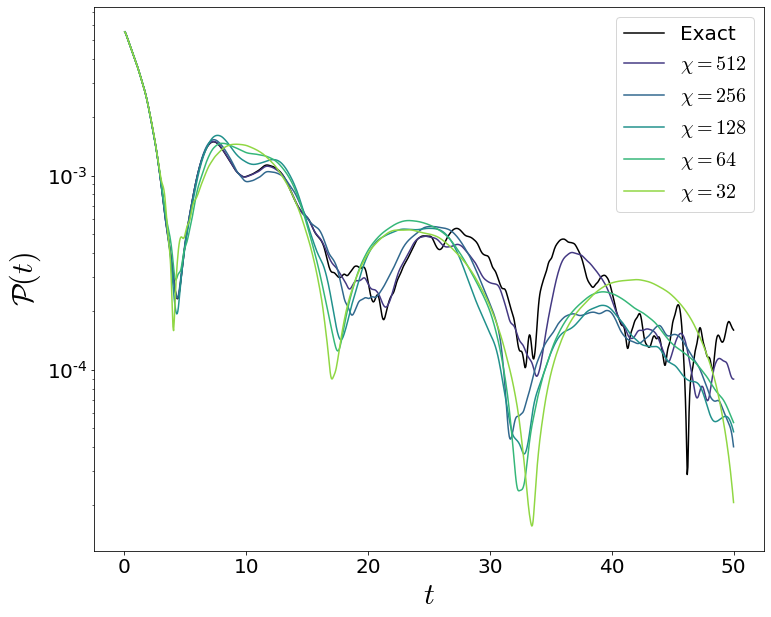}
\centering
\caption{The bipartite (the half-cut of the system) MPO entanglement entropy $\mathcal{S}(t)$ for different maximum bond dimensions for a small system size ($N = 10$) and $\lambda = 0.0$ (Upper figure). The distance function $\mathcal{P}(t)$ of the same model and parameters (Lower figure). }
\label{fig:eeunitary}
\end{figure}

The truncation of the MPDO associated with the finite bond dimension certainly affects the dynamics in the unitary evolution setting. As shown in the top of Fig.~\ref{fig:eeunitary}, while the exact result without truncation ($\chi = 4^5 = 1024$ for $N = 10$) for $\mathcal{S}(t)$ increases essentially monotonically, the corresponding plots with truncated bond dimension show a peak at the early time followed by a slow decay to a lower value than the true value. Thus, it is hard to track the exact dynamics of the system in this regime with moderate bond dimension.

Nevertheless, the diffusive character of the system can still be revealed with a modest bond dimension. The diffusion constant in this case can be extracted from a `distance' function,
\begin{equation}
    \mathcal{P}(t) = \sqrt{ \sum_{i = 1}^{N} \left ( \langle \sigma_i^z (t) \rangle - \frac{S_z}{N} \right )^2 },
\end{equation}
where $S_z$ is the total magnetization of the fully relaxed state. 
At a long enough time, the distance function relates to the diffusion constant by $\mathcal{P}(t) \propto e^{-\pi^2 t D / N^2}$.
As shown in the bottom of Fig.~\ref{fig:eeunitary}, $\mathcal{P}(t)$ resembles the exact result much better than $\mathcal{S}(t)$ within the same bond dimension range.
This indicates that transport properties and their exponents extracted from unitary evolution is reliable while not perfect.

\begin{figure}[h]
\includegraphics[width=0.9\linewidth]{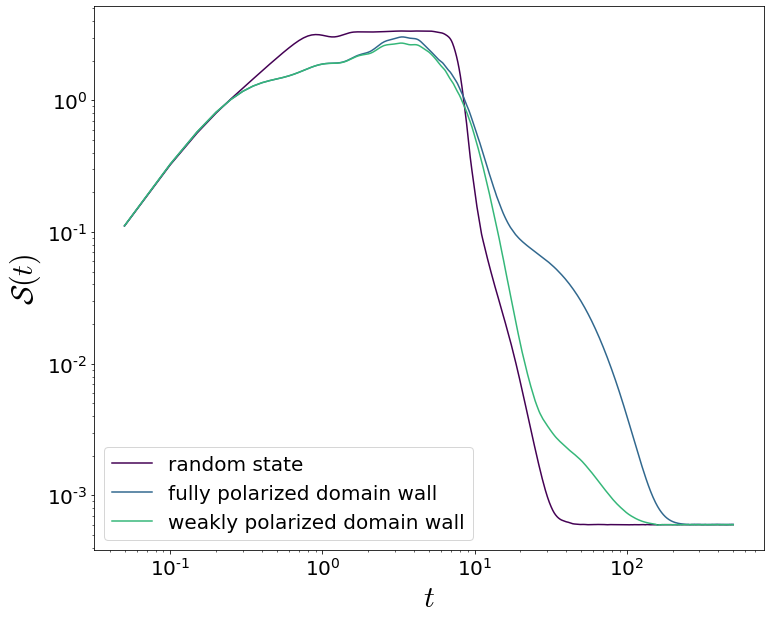}
\centering
\caption{The bipartite (the half-cut of the system) MPO entanglement entropy $\mathcal{S}(t)$ obtained from different initial superket states for the Lindbladian time evolution and system size ($N = 24$) and $\lambda = 0.0$.}
\label{fig:eelind}
\end{figure}

For the Lindbladian setting, in contrast, there may be convergence issue for different initial states we choose. 
We again turn to the calculation of $\mathcal{S}(t)$ with a number of initial states and find $\mathcal{S}(t)$ of all randomly prepared initial states appropriately converge to the same value in practically accessible time and bond dimension [Fig. ~\ref{fig:eelind}].
Therefore, we confirm that the NESS with correct entanglement structure can be efficiently obtained by our method.

\end{document}